\documentclass[%
 aip,
 amsmath,amssymb,
 reprint,%
]{revtex4-1}

\usepackage{graphicx}
\usepackage{dcolumn}
\usepackage{bm}

\usepackage[utf8]{inputenc}
\usepackage[T1]{fontenc}
\usepackage{mathptmx}
\usepackage{etoolbox}

\makeatletter
\def\@email#1#2{%
 \endgroup
 \patchcmd{\titleblock@produce}
  {\frontmatter@RRAPformat}
  {\frontmatter@RRAPformat{\produce@RRAP{*#1\href{mailto:#2}{#2}}}\frontmatter@RRAPformat}
  {}{}
}%
\makeatother
\begin{document}

\preprint{AIP/123-QED}

\title[]{Programming universal unitary transformations on a general-purpose silicon photonics platform}
\author{José Roberto Rausell-Campo}
\email{joraucam@upv.es}
 \affiliation{Photonics Research Lab, iTEAM. Universitat Politècnica de Valencia}
\author{Daniel Pérez-López}%
\affiliation{ 
IPronics Programmable Photonics S.L.
}%

\author{José Capmany Francoy}
\affiliation{%
Photonics Research Lab, iTEAM. Universitat Politècnica de Valencia
}%

\date{\today}

\begin{abstract}
General-purpose programmable photonic processors provide a versatile platform for integrating diverse functionalities on a single chip. Leveraging a two-dimensional hexagonal waveguide mesh of Mach-Zehnder interferometers, these systems have demonstrated significant potential in microwave photonics applications. Additionally, they are a promising platform for creating unitary linear transformations, which are key elements in quantum computing and photonic neural networks. However, a general procedure for implementing these transformations on such systems has not been established yet. This work demonstrates the programming of universal unitary transformations on a general-purpose programmable photonic circuit with a hexagonal topology. We detail the steps to split the light on-chip, demonstrate that an equivalent structure to the Mach-Zehnder interferometer with one internal and one external phase shifter can be built in the hexagonal mesh, and program both the triangular and rectangular architectures for matrix multiplication. We recalibrate the system to account for passive phase deviations. Experimental programming of 3x3 and 4x4 random unitary matrices yields fidelities > 98$\%$ and bit precisions over 5 bits. To the best of our knowledge, this is the first time that random unitary matrices are demonstrated on a general-purpose photonic processor and pave the way for the implementation of programmable photonic circuits in optical computing and signal processing systems.
\end{abstract}

\maketitle

\section{\label{sec:introduction}Introduction}
In the previous years, general-purpose programmable photonic integrated processors have emerged as a promising platform for the inclusion of a variety of functionalities on a single platform, similar to field-programmable gate arrays in electronics. \cite{Bogaerts2020, Perez2018, Perez2017} These systems allow software-controlled manipulation of light paths across a 2D waveguide mesh using their 2x2 building blocks, known as programmable unit cells (PUCs). PUCs are typically built from symmetric Mach-Zehnder interferometers with a phase shifter on each arm. The literature reports various waveguide mesh topologies, with triangular, rectangular, and hexagonal being the most common. \cite{Perez-Lopez:19, Zhuang:15, Perez2018}

One of the main application areas of programmable photonic circuits is for microwave photonics systems, \cite{Marpaung2019} where RF signals are processed in the optical domain and thus, benefiting from the increased bandwidth and reduced latency when compared with its electronic counterparts.\cite{7805240} Recently, researchers presented a general-purpose programmable silicon photonic circuit with a hexagonal topology. \cite{Perez-Lopez2024} They demonstrated the complete system, including the photonic circuit, electronic drivers, and software layer. Their work experimentally showcased 12 microwave photonic functionalities, highlighting the versatility of these devices.

Another area of interest is the implementation of unitary linear transformations. Application-specific photonic circuits have demonstrated capabilities for performing unitary matrix multiplications using coherent structures that combine beam splitters and phase shifters within a planar architecture. \cite{Harris:18} The basic building blocks, Mach-Zehnder interferometers with internal and external phase shifters, can be connected in various arrangements. Rectangular (Clements) \cite{Clements2016} and triangular (Reck) \cite{Reck1994} topologies are the two most common approaches. An alternative using a PUC as a building block has also been proposed to reduce area by eliminating the external phase shifter. \cite{10.1063/5.0053421} These coherent circuits inherently apply unitary transformations, which are crucial for applications like quantum computing. \cite{Harris2017, doi:10.1126/science.aab3642, Wang2020, Taballione2023modeuniversal} Additionally, unitary systems can be used to unscramble the mixing of optical signals traveling through multimode fibers \cite{Annoni2017, Choutagunta2020, Zhou2020} or free-space optical communication channels. \cite{SeyedinNavadeh2024} Finally, general matrix multiplications can be achieved using SVD decomposition, where two unitary matrices and a diagonal matrix are concatenated. \cite{Miller:13} Photonic general matrix multiplications have shown promise for deep learning, \cite{Shen2017, doi:10.1126/science.ade8450, Zhang2021} unconventional computing approaches, \cite{Prabhu2020, Roques-Carmes2020} and RF signal separation. \cite{Zhang2023, 10092945, Zhang2024} 

A general procedure for the implementation of linear unitary transformations on general-purpose photonic platforms has not been reported yet. While Perez et al. (2017) demonstrated programming a rectangular interferometer within a 7-cell hexagonal waveguide mesh, their experimental validation was limited to implementing permutation matrices (unitary transformations with absolute values of elements equal to 0 or 1). This was a consequence of two reasons: First, due to the mesh size, the input vector to be multiplied by the matrix has to be encoded externally and thus, compromising coherence and rendering complex matrix-vector multiplications infeasible. Second, it is known that due to fabrication imperfections, the initial state of the phase shifters is not 0 rad and a calibration of that passive phase must be performed.\cite{Bandyopadhyay2022} The required steps to perform this calibration were not provided nor theoretically or experimentally.

In the following work, we demonstrate how to program arbitrary unitary transformations on a general-purpose programmable photonic chip with a hexagonal topology. We utilize the commercially available Smartlight processor from iPronics Programmable Photonics. \cite{Perez-Lopez2024, iPronics} First, we explain the necessary steps for splitting light to maintain on-chip coherence. Second, we show how to replicate the transfer function of a Mach-Zehnder interferometer with internal and external phase shifters using two programmable unit cells (PUCs). We then demonstrate that both triangular and rectangular architectures can be programmed on this platform and present the recalibration process to measure the passive state of the phase shifters. We experimentally program 4x4 and 3x3 random unitary matrices. We calculate their fidelity and bit precision and perform random complex matrix-vector multiplications. Finally, we show how these general-purpose processor can be a promising platform for photonic neural networks and quantum circuits by solving two classification tasks with simulated feedforward neural networks and by programming a set of quantum logic gates.

\section{\label{general}Hexagonal Programmable Photonic Circuits}
\begin{figure}
    \centering
    \includegraphics[width=1.0\linewidth]{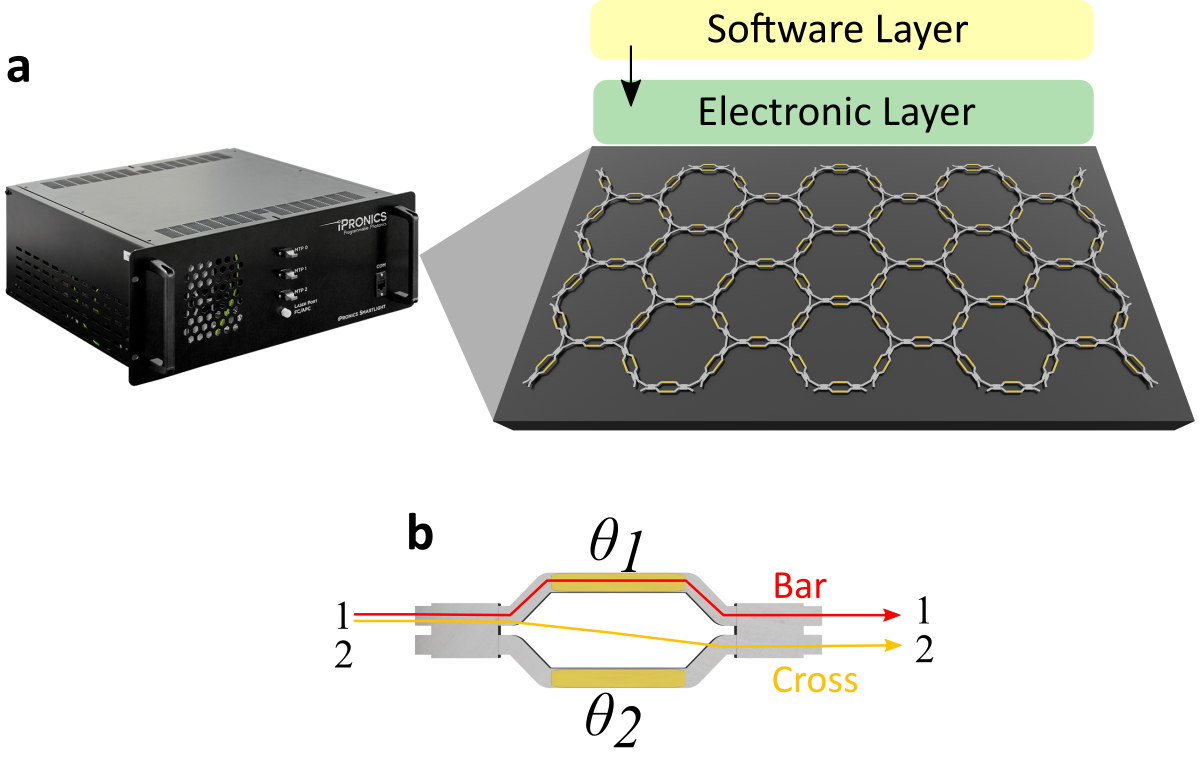}
    \caption{\textbf{a} Smarlight processor from IPronics. It combines an hexagonal waveguide mesh with an electronic and software layer, and \textbf{b} Programmable unit cell (PUC) of the hexagonal mesh. It consist of two internal phase shifters $\theta_{1}$ and $\theta_{2}$.}
    \label{fig:smartlight}
\end{figure}

The hexagonal topology have appeared to be the most promising topology for general-purpose programmable processors, see Fig. \ref{fig:smartlight}\textbf{a}. In this section, we provide their fundamental working principle and calibration requirements. 

\subsection{Programmable unit cell}
The basic building block of the hexagonal mesh is the programmable unit cell which comprises a mach zehnder interferometer with a thermo-optic phase shifter in each of the top and bottom arms. A picture of the PUC is shown in Fig. \ref{fig:smartlight}\textbf{b}. The transfer function of the PUC can be expressed as:

\newcommand{\FF}{\vphantom{\frac{A_{i,1}y^2}{y^2}}}
\begin{gather}
\renewcommand{\arraystretch}{1.3}
    ie^{i\frac{\theta_{1} + \theta_{2}}{2}} 
    \begin{pmatrix}
        sin(\frac{\theta_{1} - \theta_{2}}{2}) & cos(\frac{\theta_{1} - \theta_{2}}{2})\\
        cos(\frac{\theta_{1} - \theta_{2}}{2}) & -sin(\frac{\theta_{1} - \theta_{2}}{2})
    \end{pmatrix}
    \label{eq:puc1}\\
    = ie^{i\phi} 
    \begin{pmatrix}
        \sqrt{1 - k} & \sqrt{k}\\
        \sqrt{k} & \sqrt{1 - k}
    \end{pmatrix}
    \label{eq:puc2}
\end{gather}

The PUC enables the control of the coupling ratio (\textit{k}) between the input 1 and output 2, as well as the phase shift ($\phi$) experienced by the optical signal. This control is achieved by adjusting the relative phase applied to the top and bottom arms. In the bar state ($k$ = 0), light entering port 1 exits from port 1, and light entering port 2 exits from port 2. Conversely, in the cross state ($k$ = 1), light entering port 1 exits from port 2, and light entering port 2 exits from port 1. 

\subsection{Calibration}\label{sec:calibration}
The calibration process aims to find the relationship between the applied current on the thermo-optic actuator and the resulting phase shift, which can be expressed by the following equation:
\begin{equation}
    \theta = \theta_{0} + \alpha I^{2}
    \label{eq:calibration}
\end{equation}
where $\alpha$ is the proportionality constant between the current and the phase term, and $\theta_{0}$ is the passive phase that appears as a consequence of the fabrication imperfections, creating a difference in the group index between the top and bottom waveguides of the MZI. In an ideal system, if we set $\theta_{1} = \theta_{2}$ = 0 in (\ref{eq:puc1}), the PUC is in $cross$ state. However, variations in the passive phase across different PUCs will cause the initial state to deviate from the ideal and become a random one between $bar$ and $cross$. The calibration of each individual PUC in a hexagonal waveguide mesh requires the use of an optimization and an auto-routing algorithm. \cite{9975587, Lopez:20} Once calibrated, we can define the coupling state of each PUC, enabling functionalities like beam splitters, optical interconnects, or filters. However, for applications requiring coherence, such as unitary matrix multiplications, calibration of individual phase shifters is not sufficient. We also need to compensate for the difference in phase gained by light traversing different paths with the same number of PUCs. This phase deviation, also due to fabrication imperfections but fixed for each defined path, can be treated similarly to the passive phase of individual PUCs. In the following section, we will demonstrate how to apply this re-calibration for matrix multiplications with rectangular and triangular topologies. The same procedure can be extended to other coherent structures.

\subsection{Smarlight Processor}
Regarding the experimental demonstration of our work, we use the commercially available Smartlight photonic processor. An image of the full system with a schematic of the integrated circuit is shown in Fig. \ref{fig:smartlight}\textbf{a}. The processor comprises 17 hexagonal cells for a total of 72 programmable unit cells (PUCs). Each PUC has an insertion loss of 0.48 dB and a power consumption of 1.3mW/$\pi$. Light input and output can be done using any of the 28 available optical ports with a fiber-array that introduces a 3-dB insertion loss per facet. Each of the 28 I/O ports features an opto-electronic monitoring unit with on-chip photodetectors, enabling optical power measurement without external units. The system also includes the electronic circuitry and the software layer used to control each PUC. Further details on the device's performance and elements are available in Ref. \cite{Perez-Lopez2024}.

\section{Unitary transformations on a photonic processor}
In the following section, we detail the configuration steps required within the photonic processor to achieve vector-matrix multiplications.

\subsection{Splitter tree}\label{sec:splitter}
To ensure coherence throughout the operation, we employed a continuous-wave (CW) laser. The light was then split into a number of paths equal to the number of matrix inputs. For this work, we focused on matrices of sizes 3x3 and 4x4. Conventionally, splitter trees for dividing light equally are built using cascaded stages of multi-mode interferometers (MMIs) or directional couplers with a fixed 50:50 coupling ratio. However, in a general-purpose programmable photonic processor, different paths may experience varying optical losses due to the differing number of PUCs traversed by light on each path.  To address this challenge, we implemented a correction to the coupling ratio programmed within the PUCs. This correction takes into account the measured insertion losses of each PUC. Assuming that the coupling ratio of light to a path with fewer PUCs is denoted by \textit{k}, to achieve equal light intensity on both output paths, the following relationship applies:
\begin{equation}
    (1-k)IL^{N_{P}} = k
    \label{eq:splitter}
\end{equation}
where $IL$ are the linear insertion losses of the PUC and $N_{P}$ is the difference in number of PUCs between the two paths. Solving $k$ we obtain:
\begin{equation}
    k = \frac{1}{1 + \frac{1}{IL^{N_P}}}
    \label{eq:coupling}
\end{equation}
\begin{figure}
    \centering
    \includegraphics[width=0.85\linewidth]{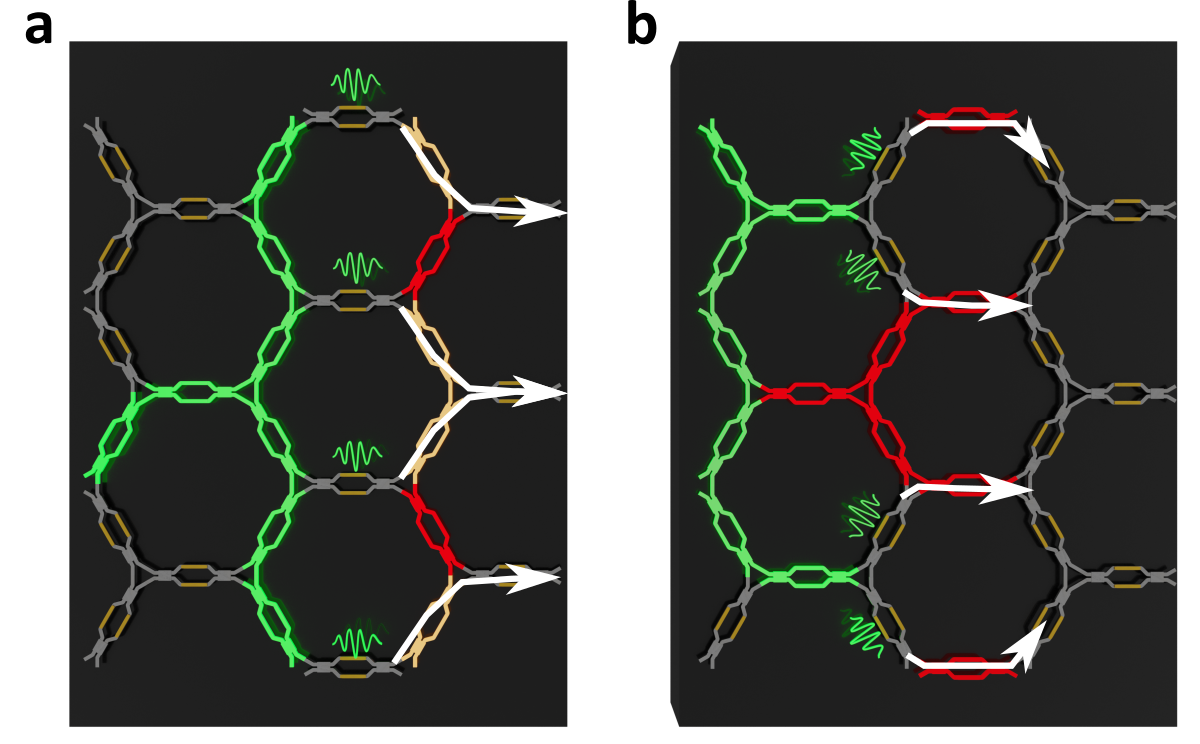}
    \caption{\textbf{a} 1x4 Symmetric splitter tree, and \textbf{b} 1x4 Non-symmetric splitter tree.}
    \label{fig:splitter}
\end{figure}

Two possible implementations for a 1x4 splitter tree are depicted using green light in Fig. \ref{fig:splitter}. Grey PUCs represent a random state except those with a green wave on them that refer to the PUCs used for input vector encoding. Data encoding is explained in the following section, but it is necessary to state that only light going out one of the two output ports of the encoding PUC is used for the matrix multiplication. The light exiting the other port is routed out of the processor before the multiplication stage. To do so, we set some PUCs in bar state (red) and others in cross (yellow) depending on the programmed splitter tree. White arrows indicate the travel path of the encoded input vector to be multiplied by the matrix.

The symmetric splitter tree in Fig. \ref{fig:splitter}\textbf{a} requires three columns of PUCs. First, the input light is divided using a 50:50 coupling. For the second division, we need to be aware that one path will undergo two more PUCs than the other (compare the path difference between the two upper outputs of the splitter tree).  Considering the insertion losses of the Smartlight processor and equation (\ref{eq:coupling}), the coupling ratio in this second stage needs to be adjusted to 56:44.

The non-symmetric splitter tree in Fig. \ref{fig:splitter}\textbf{b} is more compact as it only requires two columns. On the other hand, ligth divided in the first stage undergoes 4 more PUCs in the lower branch than in the upper. Following the path length compensation equation (\ref{eq:coupling}), the coupling ratio must be 61:39.

\subsection{Encoding}
Input vector encoding is performed using an array of PUCs, see the PUCs with a green wave on them in Fig. \ref{fig:splitter}. For the case of the symmetric splitter tree we used the $bar$ port to encode the data while for the non-symmetric splitter we used the $cross$ port.  In the general case, the input vector is an array of complex numbers. We encode its modulus in the amplitude of the optical signal and its angle (phase) using the term $\phi$. Assuming the light intensity at each input port is normalized to 1, the encoded modulus becomes equal to $\sqrt{k}$ or $\sqrt{1-k}$, depending on whether we are using the $cross$ or $bar$ ports. Focusing on the $cross$ encoding and from equations (\ref{eq:puc1}) - (\ref{eq:puc2}) we find that: 
\begin{gather}
    cos(\frac{\theta_{1} - \theta_{2}}{2}) = \sqrt{k} \\
    \Delta\Theta = \theta_{1} - \theta_{2} = 2arccos(\sqrt{k})
\end{gather} 
Substituting this expression in the phase term of (\ref{eq:puc1}) we obtain
\begin{equation}
    \phi = \frac{\theta_{1} + \theta_{2}}{2} = \theta_{2} + arccos(\sqrt{k})
\end{equation}
and finally
\begin{gather}
    \theta_{1} = \phi + arccos(\sqrt{k})\label{eq:angle1} \\
    \theta_{2} = \phi - arccos(\sqrt{k})
    \label{eq:angle2}
\end{gather}
For the $bar$ encoding the $arccos(\sqrt{k})$ term in equations (\ref{eq:angle1}) - (\ref{eq:angle2}) is substituted by $arcsin(\sqrt{1-k})$.

\subsection{Building block equivalence}
\begin{figure}
    \centering
    \includegraphics[width=0.65\linewidth]{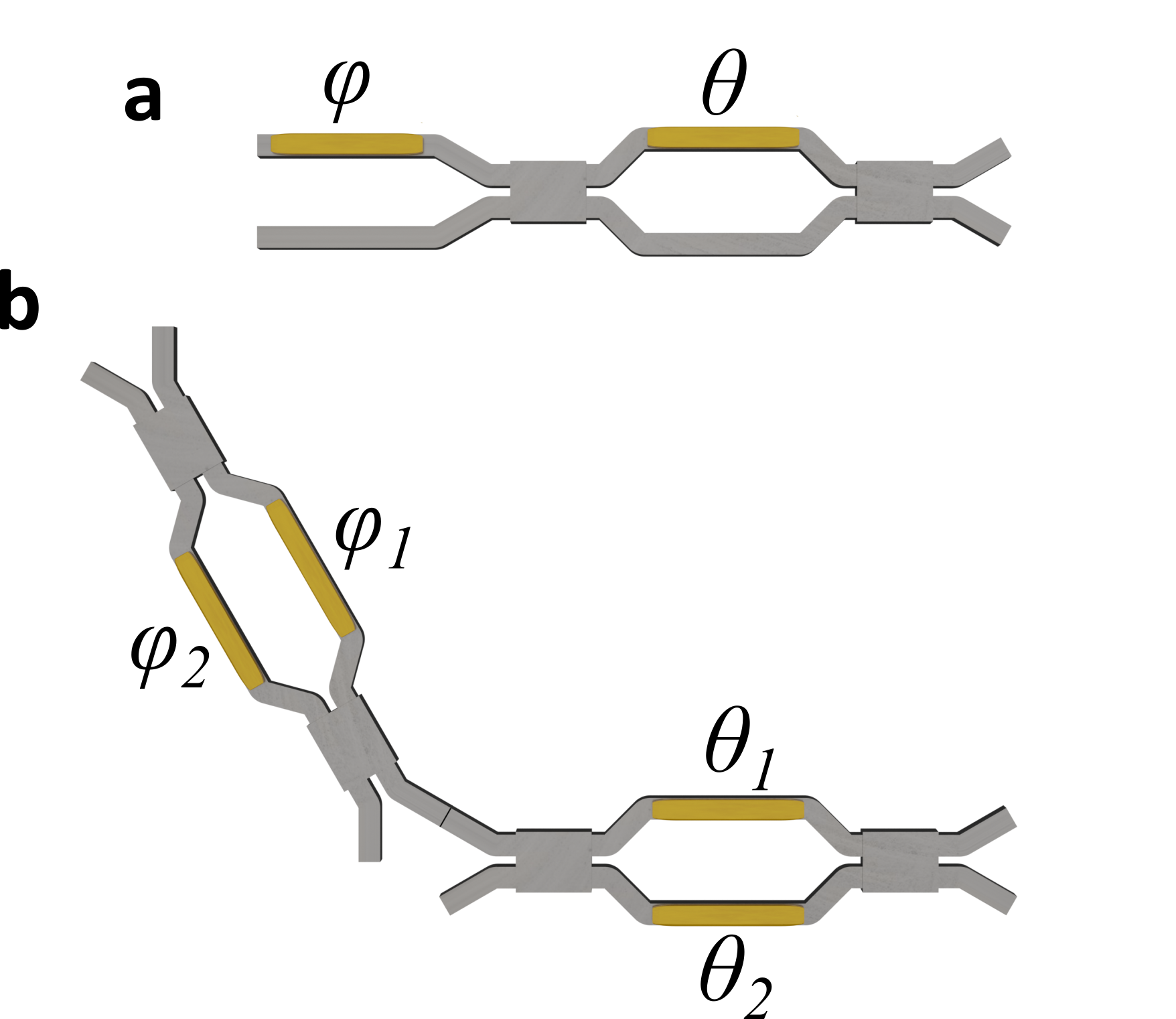}
        \caption{\textbf{a} The standard building block consists of an MZI with one external and one internal phase shifter, \textbf{b} equivalent system that concatenates two PUCs with $\theta_{2}$ = 0 and $\phi_{1}$ = $\phi_{2}$.}
    \label{fig:bbs}
\end{figure}

To program arbitrary linear interferometers on our programmable photonic chip, we first establish the equivalence between the building blocks used in each system. As mentioned earlier, the programmable unit cell (PUC) serves as the fundamental building block of the general-purpose photonic processor. In contrast, standard decomposition algorithms for linear interferometers typically assume a Mach-Zehnder interferometer (MZI) with a single internal and a single external phase shifter as the building block, see Fig. \ref{fig:bbs}\textbf{a}. This MZI has the following transfer function:

\begin{equation}
\renewcommand{\arraystretch}{1.3}
    ie^{i\frac{\theta}{2}} 
    \begin{pmatrix}
        e^{i\phi}sin\frac{\theta}{2} & e^{i\phi}cos\frac{\theta}{2}\\
        cos\frac{\theta}{2} & -sin\frac{\theta}{2}
    \end{pmatrix}
    \label{eq:bb1}
\end{equation}

We can show that if two PUCs are concatenated as in Fig. \ref{fig:bbs}\textbf{b} the same transfer function is obtained. First, phases of the left PUC must be equal, $\phi_{1}$ = $\phi_{2}$, and according to (\ref{eq:puc1}) the transfer function is:
\begin{equation}
\renewcommand{\arraystretch}{1.3}
    ie^{i\phi} 
    \begin{pmatrix}
        0 & 1\\
        1 & 0
    \end{pmatrix}
    \label{eq:ps}
\end{equation}
and the PUC is in cross state, keeping the amplitude constant and adding a phase of $\phi + \frac{\pi}{2}$, which is equivalent to an external phase shifter. Then, the second PUC will be tuned maintaining $\theta_{2}$ = 0. The final linear transformation of the system is consequently the same as in (\ref{eq:bb1}).

\subsection{Matrix Architecture}
\begin{figure}
    \centering
    \includegraphics[width=1.0\linewidth]{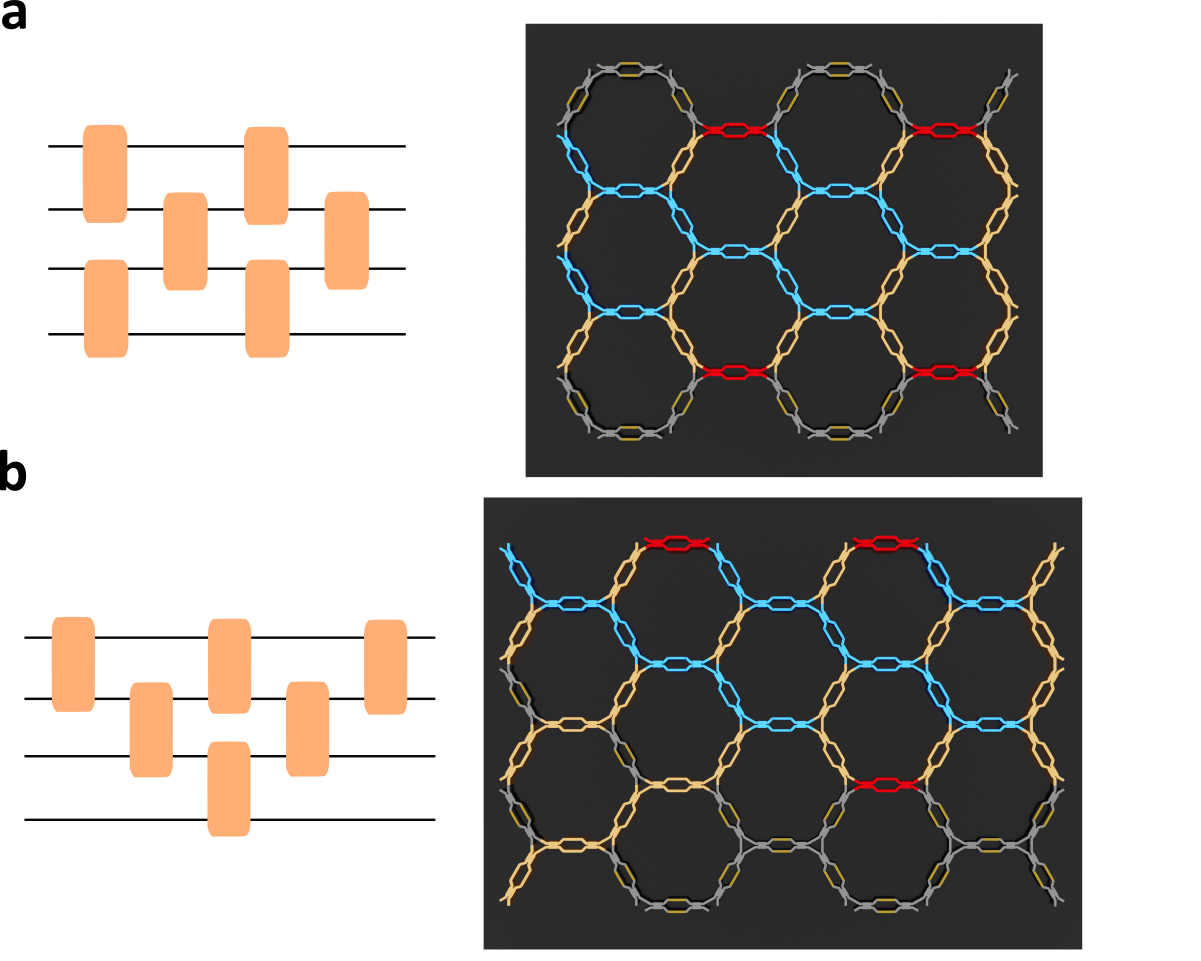}
    \caption{\textbf{a} Schematic of the Clements topology with its translation to the hexagonal mesh, and \textbf{b} schematic of the Reck topology with its translation to the hexagonal mesh.}
    \label{fig:matrices}
\end{figure}

We have established that two PUCs can be configured to function equivalently to a Mach-Zehnder interferometer (MZI) with one internal and one external phase shifter, the fundamental building block in many coherent integrated processors. This equivalence allows us to directly translate previously proposed architectures for unitary matrix multiplications on photonic circuits to our general-purpose processor. In our case, we show how to implement the rectangular (Clements) and triangular (Reck) architecture. In Fig. \ref{fig:matrices}, we show the distribution of the building blocks (orange boxes) in the Clements (Fig. \ref{fig:matrices}\textbf{a}) and in the Reck (Fig. \ref{fig:matrices}\textbf{b}) architecture. The blue PUCs represent the equivalent building blocks within the hexagonal mesh. PUCs programmed in cross and bar states ensure that light propagates in a single direction and interacts with other paths only at tunable elements.
The Reck architecture necessitates an additional column of horizontal PUCs compared to the Clements topology. As a consequence, in our experiment the symmetric splitter presented in Section \ref{sec:splitter} is not compatible with the Reck architecture using the current mesh size of the Smartlight processor and we will use the non-symmetric version. 

\subsection{Phase Calibration}
\begin{figure}
    \centering
    \includegraphics[width=1.0\linewidth]{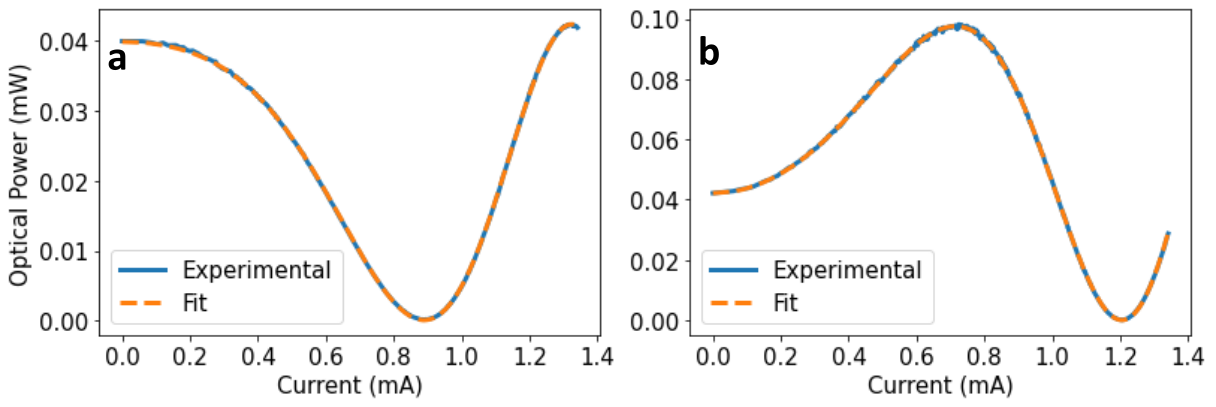}
    \caption{Optical response of two META-MZI and their fitting curve.}
    \label{fig:calibration}
\end{figure}

As discussed in Section \ref{sec:calibration}, the standard calibration method for programmable hexagonal meshes is insufficient for implementing coherent transformations. While it can measure and compensate for the passive phase difference between the top and bottom waveguides of a PUC, it doesn't account for phase variations arising from different connections between PUCs in a programmed coherent architecture. These additional phase differences can also be treated as a passive phase inherent to the chosen architecture. To address this challenge, we propose a re-calibration procedure that focuses on measuring these inherent passive phase offsets within the architecture. The procedure assumes that the $\alpha$ term in equation (\ref{eq:calibration}) is still valid and only the $\theta_{0}$ term should be measure. These offsets will be incorporated as the passive phase of each PUC acting as a phase shifter within the mesh. The re-calibration procedure is adapted from methods used for single-phase actuator calibration in application-specific integrated circuits (ASPICs). \cite{Prabhu2020, Lin:24, Alexiev2021, Pentangelo} This method relies on constructing a temporary interferometer using single PUCs. In the ASPIC approach, this structure, also known as a META-MZI, is built using two PUCs configured as 50:50 couplers with all internal PUCs set to the bar state. However, in our general-purpose processor, we have additional PUCs acting as waveguides that are set to the cross state. Once the system is defined, we perform a two-step characterization process for each PUC acting as a phase shifter. In the first step, we sweep the current applied to the top arm while keeping the bottom arm current at zero. We measure the output power at the cross port and identify the current corresponding to the maximum power output. This ensures the PUC is in the cross state. In the second step, we sweep the current of both the top and bottom arms simultaneously. We add the previously measured optimal current as an offset to the top arm current during this sweep. This effectively modifies the phase of the optical signal while maintaining the coupling ratio of the PUC. This phase modification within the created META-MZI results in an interferometric pattern as shown in Fig. \ref{fig:calibration}(\textbf{a-b}). This measurement can be fitted using equations (\ref{eq:puc1}) and (\ref{eq:calibration}) to extract the  passive phase offset of the phase shifter under test.

\begin{figure}
    \centering
    \includegraphics[width=0.75\linewidth]{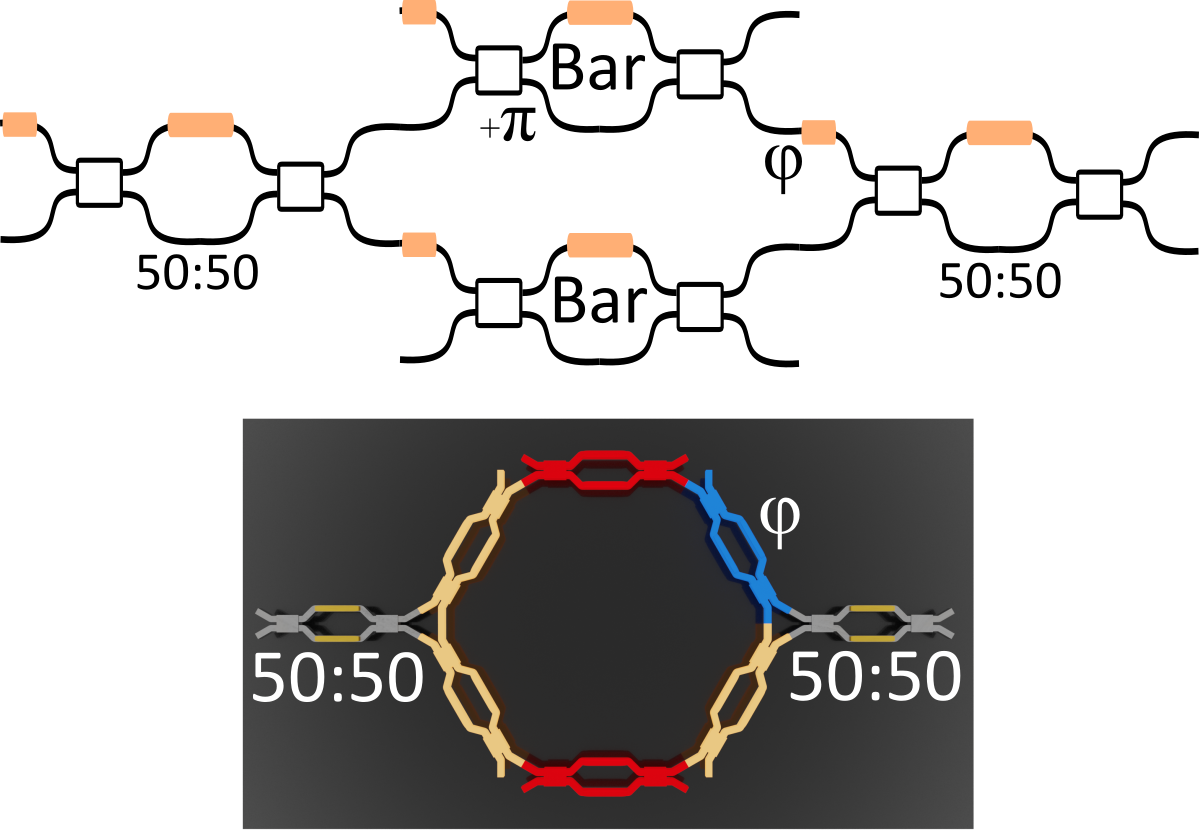}
    \caption{META-MZI method for the characterization of the passive phase of the phase shifters and its translation to the hexagonal mesh.}
    \label{fig:meta}
\end{figure}

\section{Experiments and results}
\begin{figure*}
\centering
\includegraphics[width=0.95\textwidth]{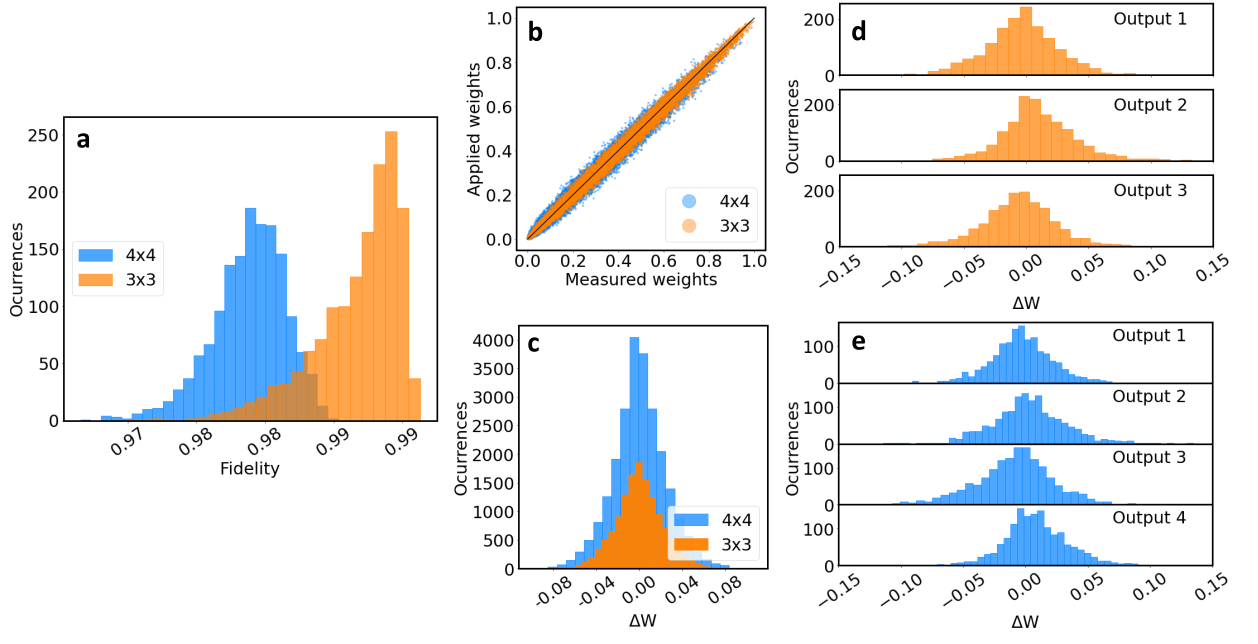}
\caption{Results of 1500 random unitary matrices using the Clements architecture with a size of 3x3 (blue) and 4x4 (orange). \textbf{a} Measured fidelity, \textbf{b} comparison between the ideal applied weights and the measured weights. The black line represents the scenario where there are no errors, \textbf{c} difference between the applied and measured weights and \textbf{d-e} difference between the measured and ideal result of the multiplication of the unitary matrix by a random vector.}
\label{fig:clements_meas}
\end{figure*}

\begin{figure*}
\centering
\includegraphics[width=0.95\textwidth]{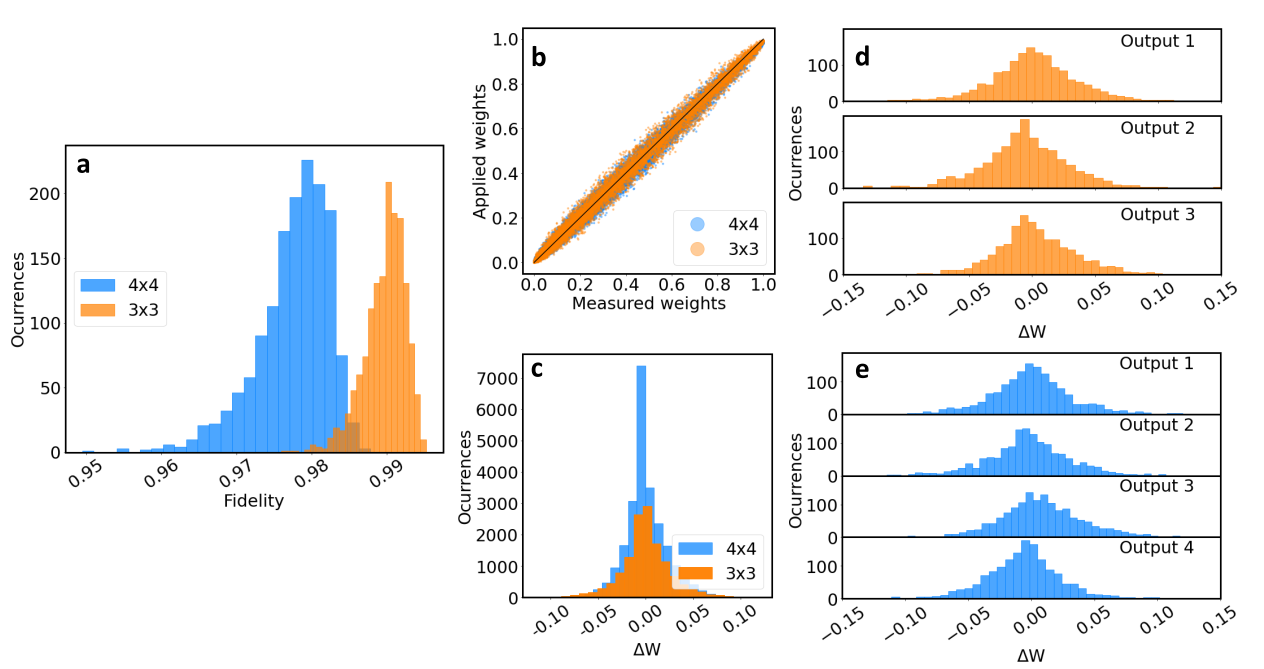}
\caption{Results of 1500 random unitary matrices using the Reck architecture with a size of 3x3 (blue) and 4x4 (orange). \textbf{a} Measured fidelity, \textbf{b} comparison between the ideal applied weights and the measured weights. The black line represents the scenario where there are no errors, \textbf{c} difference between the applied and measured weights and \textbf{d-e} difference between the measured and ideal result of the multiplication of the unitary matrix by a random vector.}
\label{fig:reck_meas}
\end{figure*}

Following re-calibration of the hexagonal mesh, we proceeded to encode unitary matrices on the Smartlight processor. All measurements utilized the same 1550 nm, 10 dBm output power laser source employed for calibration. We programmed the Clements and the Reck architecture on the photonic processor using the aforementioned scheme. For each architecture, we tested two matrix sizes: 3x3 and 4x4. The translation between the unitary matrix weights and the phases that are needed to apply was carried out following the decomposition algorithms provided in Refs. \cite{Clements2016, Reck1994} The decomposition of the 3x3 matrices results in the same building block arrangement for both architectures. To differentiate the implementations, we positioned the matrix multipliers in different regions of the mesh. The Clements architecture employed the symmetric splitter tree, while the non-symmetric splitter tree was used with the Reck architecture. During the experiments, for each combination of architecture and matrix size, we generated 1500 random unitary matrices. These matrices were decomposed into the required phase shifts and programmed onto the photonic integrated chip. We measured three figures of merit. 

First, we measured the fidelity of the matrices. Fidelity is a common metric in quantum research used to quantify the similarity between two unitary operations. It is calculated using the following equation:
\begin{equation}
    \mathcal{F} = \frac{Tr(|U^{\dagger}U|^{2})}{N}
    \label{eq:fidelity}
\end{equation}
where $U^{\dagger}$ is the adjoint matrix of U, $Tr$ denotes the trace of the matrix and $|.|^{2}$ if the element-wise square-modulus of the matrix-matrix multiplication. The highest fidelity (F = 1) indicates perfect similarity between the applied and desired unitary operations. In our chip, we include the splitter tree, input vector and matrix, allowing for the coherent vector-matrix multiplication. As a result, we encoded in the input vector the complex columns of the adjoint matrix one at a time to perform the matrix-matrix multiplication in (\ref{eq:fidelity}). The squared-modulus is applied on the photodetection stage. Finally, the trace of the recorded result is calculated to determine the measured fidelity.

Second, we assessed the accuracy of the programmed matrix weights. We encoded in the input vector the $i_{th}$ column of the identity matrix which results in the squared modulus of the $i_{th}$ row of the programmed matrix. We repeated this process sequentially for all columns, effectively reconstructing the entire programmed matrix. The measured data is then compared with the ideal matrix, and the r-squared coefficient of the linear regression between the two datasets is calculated to quantify the overall agreement. Additionally, we measure the difference between the measured and ideal weights, calculating the mean and standard deviation of the error. From this measurement, we can approximate the bit precision of our system using the following formula\cite{Zhang:22}:
\begin{equation}
    bit-precision = log2(\frac{max_{weight} - min_{weight}}{std(err)})
\end{equation}
where $std(err)$ is the standard deviation of the errors.

Finally, we evaluated the accuracy of on-chip matrix-vector multiplications. We sampled random complex vectors and they were multiplied by the programmed matrix within the photonic chip. The resulting outputs were measured and compared with the expected outcome of the ideal multiplication. For all measurements, the outputs were normalized by dividing each element by the total measured power. 

The fidelity results for the Clements architecture are presented in Fig. \ref{fig:clements_meas}\textbf{a}. We achieved an average fidelity of 99.2 $\pm$ 0.3 for the 3x3 matrix size and a 98.4 $\pm$ 0.3 for the 4x4 matrix size. Regarding weight accuracy, Fig. \ref{fig:clements_meas}(\textbf{b-c}) shows the comparison between the ideal and measured weights. The linear regression returns an $r^{2}$ coefficient of 0.992 and 0.984 for the 3x3 and 4x4, respectively. The standard deviation of the errors between the measured and ideal weights is 0.0215 and 0.0246, corresponding to a bit-precision of 5.5 and 5.35 bits, respectively. The errors for random vector-matrix multiplications (VMMs) are presented in Fig. \ref{fig:clements_meas}(\textbf{d-e}) and Table \ref{tab:mvm1} where we show the error occurred at each output.

\begin{table}[h]
\caption{\label{tab:mvm1} Mean and standard deviation of the error at each output of 1500 random vector-matrix multiplications using the Clements architecture.}
\begin{ruledtabular}
\begin{tabular}{ccccc}

    Size & $O_{1} (10^{-2})$ & $O_{2} (10^{-2})$ & $O_{3} (10^{-2})$ & $O_{4} (10^{-2})$ \\ \hline
    3x3 & 0.4 $\pm $3.2  & 1.0 $\pm $3.3 & 0.6 $\pm$ 3.2 & - \\ 
    4x4 & 0.3 $\pm $2.6 & 0.3 $\pm $3.1 & 0.8 $\pm$ 3.2 & 0.9 $\pm$ 2.7 \\ 
\end{tabular}
\end{ruledtabular}
\end{table}

For the Reck architecture, the measured fidelities are shown in Fig. \ref{fig:reck_meas}\textbf{a}. We obtained an average fidelity with the 3x3 and 4x4 matrices of 99.0 $\pm$ 0.3 and 97.8 $\pm$ 0.5. The comparison between the applied and measured square-modulus of the complex weights is presented in Fig. \ref{fig:reck_meas}(\textbf{b-c}). The $r^{2}$ coefficient of the linear regression is 0.993 for both matrix sizes. The standard deviation of the error between the measured and ideal scenario is 0.026 for the 3x3 matrix and 0.0216 for the 4x4 matrix, which translates in a bit-precision of 5.27 and 5.53 respectively. Errors of the random VMMs are presented in Fig. \ref{fig:reck_meas}(\textbf{d-e}). The mean and standard deviation are presented in Table \ref{tab:mvm2}. 

\begin{table}[h]
\caption{\label{tab:mvm2} Mean and standard deviation of the error at each output of 1500 random vector-matrix multiplications using the Reck architecture.}
\begin{ruledtabular}
\begin{tabular}{ccccc}

    Size & $O_{1} (10^{-2})$ & $O_{2} (10^{-2})$ & $O_{3} (10^{-2})$ & $O_{4} (10^{-2})$ \\ \hline
    3x3 & 0.2 $\pm $3.6  & 0.4 $\pm $3.6 & 0.3 $\pm$ 3.3 & - \\ 
    4x4 & 0.04 $\pm$ 3.5& 0.2 $\pm $3.4 & 0.8 $\pm$ 3.4 & 0.6 $\pm$ 3.1 \\ 
\end{tabular}
\end{ruledtabular}
\end{table}

\section{Applications}
\subsection{Photonic neural networks}
\begin{figure}
    \centering
    \includegraphics[width=1.0\linewidth]{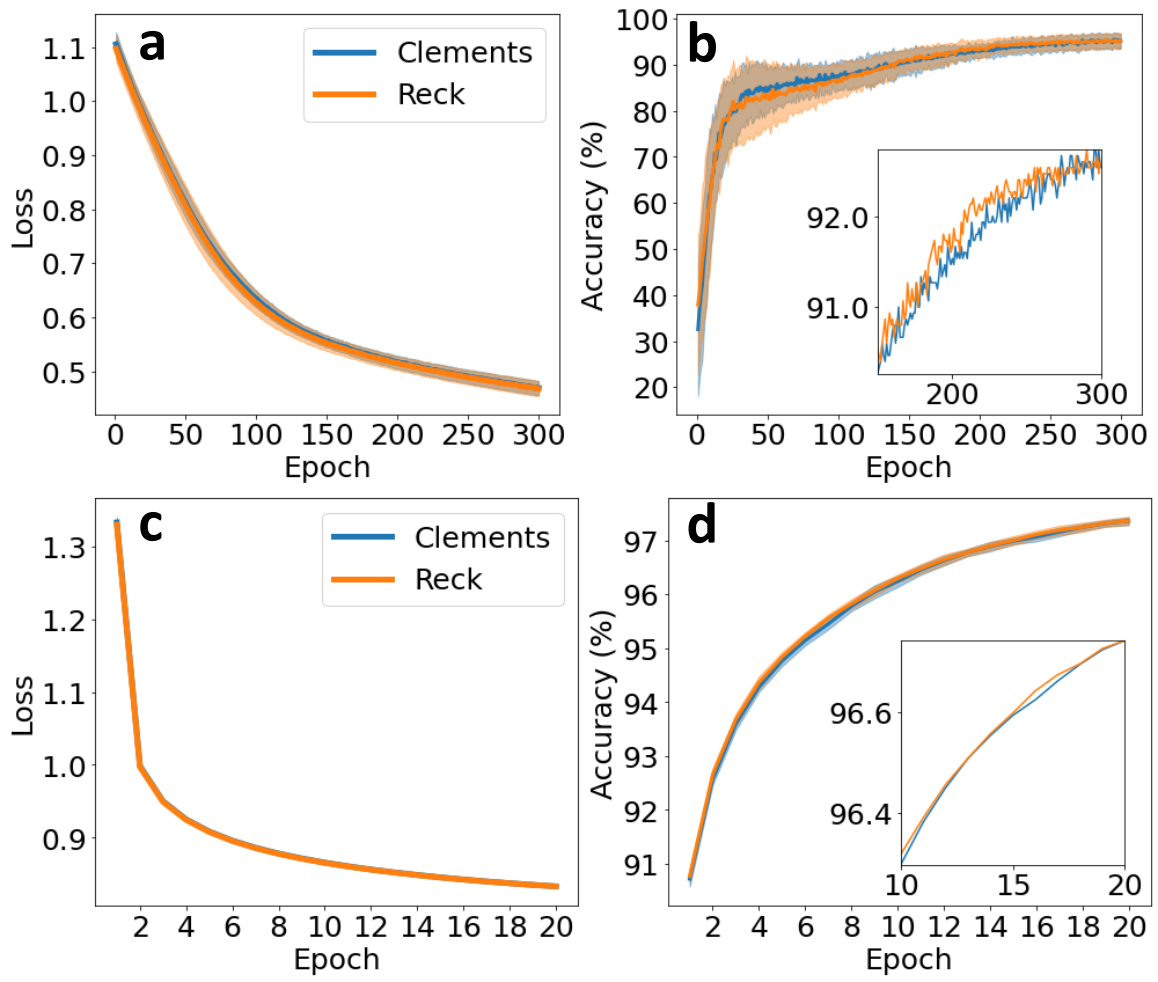}
    \caption{\textbf{a-b} Loss and accuracy on the flower classification problem using the Iris dataset, and \textbf{c-d} loss and accuracy on the handwritten classification problem using the MNIST dataset.}
    \label{fig:ml}
\end{figure}

Photonic neural networks have emerged as one of the most promising applications of photonic processors, as they can significantly reduce the power consumption and latency constraints of current electronic devices\cite{Liao2023}. The use of low-bit precision models has been proposed in the literature to reduce the computational requirements of deep learning systems\cite{ma2024era,10.5555/3295222.3295331,8578384}. To illustrate the capabilities of our programmable photonic processor, we train two benchmark models: the flower classification problem using the Iris dataset\cite{misc_iris_53}, which aims to distinguish between three types of flowers based on four input features, and the handwritten digit recognition problem using the MNIST dataset\cite{726791}. For both problems, we used feedforward neural networks with two fully connected layers and ReLU activation functions except on the output where no activation function was applied. All weights were clipped to values between -1 and 1, and after each layer, a Gaussian layer with a standard deviation corresponding to the Clements or Reck 4x4 matrix was introduced to simulate the behavior of our photonic processor. For the flower classification problem, we used 150 nodes in each layer and 300 epochs; for the handwritten recognition problem, we used 512 nodes in the first layer, 256 in the second, and 20 epochs. The network was optimized using the stochastic gradient descent algorithm and cross-entropy loss. All the models were trained 25 times. In Fig. \ref{fig:ml}\textbf{a-b}, we present the results for the Iris dataset. We show the mean training losses and the mean test accuracies during training (solid lines) and the confidence band to show the variability of the different models as a consequence of the photonic processor precision. We achieve a mean final accuracy of 95.2 $\pm$ 1.6 for both Clements and Reck architectures. The results for the MNIST dataset are presented in Fig. \ref{fig:ml}\textbf{c-d}. The final obtained accuracy is 97.358 $\pm$ 0.072 for the Clements architecture and 97.358 $\pm$ 0.076 for the Reck architecture.

\subsection{Quantum Gates}
\begin{figure}
    \centering
    \includegraphics[width=0.85\linewidth]{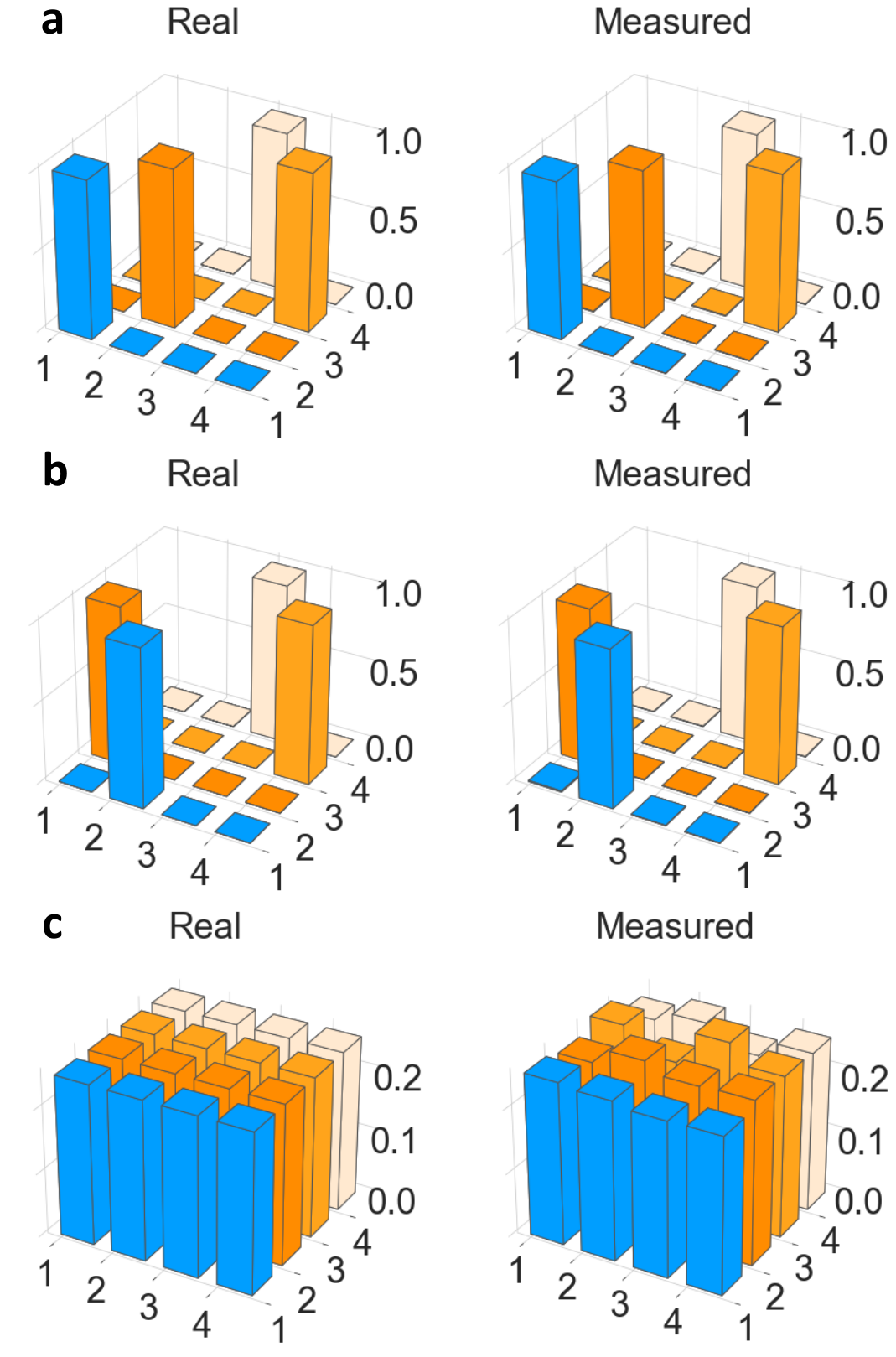}
    \caption{\textbf{a} CNOT gate, \textbf{b} Pauli Y gate and \textbf{c} Hadamard gate.}
    \label{fig:quantum}
\end{figure}

Linear photonic integrated circuits have also shown promising results by providing computational advantages in quantum systems\cite{Madsen2022}. We demonstrate the potential of general-purpose programmable photonic circuits in quantum computing applications by programming a set of quantum logic gates and comparing the experimental results with the ideal matrix. We implement the CNOT gate as shown in eq. \ref{eq:CNOT}, the Pauli Y gate extended to a 4x4 matrix as presented in eq. \ref{eq:Pauli}, and the Hadamard gate as shown in \ref{eq:Hadamard}. The experimental results are illustrated in Fig. \ref{fig:quantum}\textbf{a} for the CNOT gate, Fig. \ref{fig:quantum}\textbf{b} for the Pauli Y gate, and Fig. \ref{fig:quantum}\textbf{c} for the Hadamard gate. We achieved root mean squared errors of 0.020, 0.021, and 0.035, respectively. Regarding fidelity, we achieved highly competitive results of 0.99, 0.99, and 0.97, respectively.

\begin{equation}
\renewcommand{\arraystretch}{1.3}
    \begin{pmatrix}
    1 & 0 & 0 & 0 \\
    0 & 1 & 0 & 0 \\
    0 & 0 & 0 & 1 \\
    0 & 0 & 1 & 0
\end{pmatrix}
\label{eq:CNOT}
\end{equation}

\begin{equation}
\renewcommand{\arraystretch}{1.3}
\begin{pmatrix}
    0 & -i & 0 & 0 \\
    i & 0 & 0 & 0 \\
    0 & 0 & 0 & -i \\
    0 & 0 & i & 0 
\end{pmatrix}
\label{eq:Pauli}
\end{equation}

\begin{equation}
\renewcommand{\arraystretch}{1.3}
\frac{1}{2}
\begin{pmatrix}
    1 & 1 & 1 & 1 \\
    1 & -1 & 1 & -1 \\
    1 & 1 & -1 & -1 \\
    1 & -1 & -1 & 1 
\end{pmatrix}
\label{eq:Hadamard}
\end{equation}

\section{Discussion and Conclusions}
In this study, we have experimentally demonstrated the implementation of 3x3 and 4x4 random unitary matrices on a general-purpose programmable photonic processor. This marks, to the best of our knowledge, the first instance where random unitary matrices have been programmed on these type of platforms using both the rectangular and the triangular arrangements. The success of this implementation is attributed to the adaptation of the META-MZI algorithm for the calibration of the architecture-specific passive phases and to the construction of a building block mathematically equivalent to the mach-zehnder with one internal and one external phase shifter. Our finding indicate that both the rectangular and triangular architectures give similar results in terms of performance. The small difference in fidelity and bit-precision are within the margins of error. The obtained fidelities are compatible with quantum experiments.\cite{Pentangelo, Maring2024} Moreover, 5-bit precisions have demonstrated to be sufficient for quantized deep neural networks,\cite{ma2024era,10.5555/3295222.3295331,8578384} highlighting the capability of the processor to support advanced computational tasks. To highlight these statements, we trained two neural networks using the measured experimental data to solve the Iris classification problem and the MNIST handwritten recognition problem, achieving competitive performances. Moreover, we demonstrated the capabilities of the general-purpose processors on quantum tasks by programming the CNOT, Pauli Y, and Hadamard gates, achieving RMSEs of less than 0.035 and fidelities greater than 97$\%$.

While our results are promising, it is important to acknowledge the scalability challenges associated with the system. Compared to application-specific systems, our general-purpose processor exhibits additional losses. In the splitter tree, we use PUCs instead of 3-dB MMI which present lower insertion losses in commercially available fabrication process. Furthermore, we have also demonstrated that as the size of the matrix scales we need to add extra PUCs to the splitter tree. The matrix part also introduces extra losses. We require twice the number of PUCs compared to an ASPIC. Our system presents 0.48 dB insertion loss for PUC. Reducing this number becomes crucial for large-scale implementations. PUCs based on thermo-optic phase shifters with < 0.1 dB of losses have been already demonstrated\cite{9895610}, opening the path for high-dimensional photonic linear transformers. It is also possible to increase the size of the matrix multiplication by dividing the multiplication process into lower size multiplications using a compiler.\cite{9851450}

Regarding the power consumption of the system, each PUC in the processor consumes 1.3 mW/$\pi$. The splitter tree requires 14 PUCs which is an extra average consumption of 18.2 mW if compared with an ASPIC. For the matrix stage, we not only require the tunable elements but also PUCs to be in $bar$ and $cross$ state for light routing. Moreover, each building block requires 4 phase shifters instead of the 2 used in current ASPICs. The total average consumption of the matrix is 54.6 mW and 61.1 mw for a 4x4 matrix size using the Clements and Reck architecture, respectively. These values can be reduced by using non-volatile phase change materials\cite{Wuttig2017} for the static PUCs and by using alternative building blocks for more compact and less power-consuming  architectures. \cite{10.1063/5.0053421}  

An important metric is the number of multiply-and-accumulate (MAC) operations per second that can be achieved. In our case, we used thermo-optic phase shifters with a switching speed limited to several microseconds for both vector and matrix encoding. As our system can perform complex operations, the number of MAC/s is $MAC/s = 4N^{2}DR$, where $DR$ is the data rate. Then, our 4x4 matrices can provide up to 0.64 GMAC/s. The inclusion of high-speed modulators for the input vector encoding can increase this figure to the TMAC/s range. For example, using a DR of 20 GS/s will translate into 1.28 TMAC/s. No additional changes would be necessary as high-speed photodetectors working up to 40 GHz are already integrated in the processor.

The use of a general-purpose processor presents significant advantages, particularly in terms of reducing costs associated with design and fabrication. The flexibility of these processors also allows the integration of extra functionalities such as photonic filters or delay lines, which has been already combined with linear transformations in ASPICs for different coherent applications.\cite{Nakajima2021, 10092945} Future work could explore how the integration of these systems could impact the performance and precision of the unitary transformation.

\begin{acknowledgments}
This work was supported by the H2020-ICT2019-2 Neoteric 871330 project, the European Research Council (ERC) Advanced Grant programme under grant agreement No. 101097092 (ANBIT), the ERC Starting Grant programme under grant agreement No. 101076175 (LS-Photonics Project), and the EUR2022-134023 grant funded by CIN/AEI/10.13039/501100011033 and the European Union (NextGenerationEU/ PRTR)
\end{acknowledgments}

\section*{Data Availability Statement}
The data that support the findings of this study are available from the corresponding author upon reasonable request.

\section*{References}
\bibliography{bibliography}

\end{document}